\documentclass[aps,prb,floatfix,showpacs,reprint,citeautoscript]{revtex4-2}
\usepackage{chemformula} 
\usepackage[T1]{fontenc} 
\usepackage{siunitx}
\usepackage{amssymb}
\usepackage{amsmath}
\usepackage{graphicx}
\usepackage{multirow,makecell}
\usepackage{booktabs}
\usepackage{footnote}
\usepackage[flushleft]{threeparttable}
\usepackage{comment}
\usepackage{lineno}

\usepackage{hyperref}
\hypersetup{
    colorlinks,%
    citecolor=magenta,%
    linkcolor=magenta,%
    urlcolor=magenta
}

\providecommand{\U}[1]{\protect\rule{.1in}{.1in}}

\date{\today}

\begin{document}

\title{Effects of altermagnetic order, strain and doping in RuO$_2$}

\author{Darshana Wickramaratne}
\email{darshana.k.wickramaratne.civ@us.navy.mil}
\affiliation{US Naval Research Laboratory, Washington,
D.C. 20375, USA}
\author{Marc Currie}
\affiliation{US Naval Research Laboratory, Washington,
D.C. 20375, USA}
\author{Shelby S. Fields}
\affiliation{US Naval Research Laboratory, Washington,
D.C. 20375, USA}
\author{Cory D. Cress}
\affiliation{US Naval Research Laboratory, Washington,
D.C. 20375, USA}
\author{Steven P. Bennett}
\affiliation{US Naval Research Laboratory, Washington,
D.C. 20375, USA}
\begin{abstract}
RuO$_2$, one of the most widely studied transition metal oxides, was recently predicted
to host a novel form of collinear magnetic order referred to as altermagnetism.
In this study we combine experiment (reflectance, transmittance, ellipsometry
and Raman measurements) and first-principles calculations to elucidate the potential role of
altermagnetic order, strain and doping on the optical and vibrational properties of RuO$_2$
grown on TiO$_2$ (001), (101) and (110) substrates.
Our combined experimental and theoretical results point toward a nonmagnetic ground state as the most consistent
description of the optical and vibrational properties of bulk RuO$_2$. RuO$_2$ strained to TiO$_2$ (001) remains nonmagnetic in our calculations.
Straining to TiO$_2$ (110) stabilizes ferromagnetic and altermagnetic states that are nearly degenerate and both lower in energy than the nonmagnetic state.  The relative energetic ordering of these states is highly sensitive to the level of strain and the choice of exchange-correlation approximation.
\end{abstract}

\maketitle

\section{Introduction}
Ruthenium dioxide (RuO$_2$), which belongs to a large family of transition metal oxides that is stable in the
rutile structure \cite{rogers1969crystal} has been the topic of several studies motivated in part by its potential
applications in electro-chemical energy storage \cite{balaya2003fully} and catalysis \cite{karamad2015mechanistic}.
Additional interest
has been driven by the observation of superconductivity in RuO$_2$ thin films \cite{uchida2020superconductivity,ruf2021strain}.
The combination of potential applications and novel phenomenon enabled by RuO$_2$,
 has led to a plethora of experimental studies which have shown
the electrical conductivity is metallic at
room temperature \cite{ryden1968temperature},
the optical reflectance is high in the visible range of the spectrum \cite{goel1981optical},
the temperature dependence of the specific heat and magnetic susceptibility lack any signatures
of phase transitions \cite{passenheim1969heat,ryden1970magnetic} and early measurements
of the Fermi surface \cite{graebner1976magnetothermal}
appear to be in agreement with
calculations of the nonmagnetic electronic structure \cite{mattheiss1976electronic,glassford1993electronic}.
All of these experimental results are hallmarks of a conventional paramagnetic metal.

The recent surprising prediction that RuO$_2$ hosts a collinear magnetically ordered state,
now referred to as altermagnetism (AM), \cite{smejkal2022beyond}
has justifiably spurred research that is focused on identifying and controlling AM in RuO$_2$.
An AM has spin sublattices with equal but opposite sign magnetic moments
that are coupled by symmetry operations that are not translations or inversions \cite{smejkal2022beyond}.
This manifests in finite momentum-dependent spin splitting
in the band structure (characteristic of a ferromagnet) and zero net magnetization (characteristic of an
antiferromagnet) \cite{smejkal2022emerging,hayami2019momentum,yuan2020giant,mazin2021prediction},
which promises potential applications in the field of spintronics \cite{bose2022tilted}.

While experiments on RuO$_2$ thin films \cite{fedchenko2024observation,bose2022tilted,feng2022anomalous,jeong2025metallicity,noh2025tunneling}
appear to confirm this prediction of AM in RuO$_2$, studies on bulk RuO$_2$ 
suggest that the evidence for AM order remains inconclusive and the material
 is likely non-magnetic (NM)
\cite{hiraishi2024nonmagnetic,kessler2024absence,wenzel2024fermi}.
One possible reason for these two schools of thought
is the potential role played by strain given that some of these experiments
that were interpreted as evidence of AM were conducted on thin films grown on
different substrates \cite{zhu2019anomalous,bose2022tilted}.
Fermi level shifts due to the presence of defects have also been raised as
a possible origin for AM order \cite{smolyanyuk2024fragility}.
A range of techniques have been proposed to detect AM order,
which includes but is not limited to magneto-optical measurements \cite{zhou2021crystal,rao2024tunable,adamantopoulos2024spin}
and low-energy Raman scattering measurements to identify magnon modes \cite{vsmejkal2023chiral}.

Since the experimental effort to detect AM order in RuO$_2$ relies on understanding its electronic, optical, and vibrational properties, a prosaic yet pertinent question is whether these properties are better described by assuming a non-magnetic or an altermagnetic state.
To address this, we combine optical reflectance, transmittance, ellipsometry, and Raman scattering measurements with first-principles calculations to probe the optical and vibrational properties of RuO$_2$, accounting for different crystallographic orientations, strain conditions, doping, and temperature.
Although these measurements do not directly probe broken time-reversal symmetry
they provide a reasonable test on whether the observed one-electron excitations and structural properties
are described by an electronic structure that is NM or AM.

\section{Methods}
\subsection{Growth and characterization}
RuO$_2$ thin films were grown on double-side polished TiO$_2$ substrates (MTI) with (001), (101), and (110) orientations
through a high-temperature reactive sputtering process shown previously to produce high-quality,
lattice matched single crystal films \cite{fields2024orientation}.
For this deposition, all three substrates were loaded into the AJA ATC Orion sputter system, which has a base pressure of 2$\times$10$^{-8}$ Torr,
without surface treatment immediately following removal from hermetic packaging from the vendor.
The substrates were then heated to 450 $^{\circ}$C and allowed to equilibrate in a background pressure
consisting of 4 mTorr of Ar (7.5 sccm) and O$_2$ (7.5 sccm) to remove adventitious organics.
160 minutes of deposition was conducted through application of 30 W of power (supplied by an AEI Pinnacle Pulsed DC Power Supply
set at 100\%~duty cycle) across a 2 inch pure Ru (99.5\%~, ACI alloys) target.
After, the substrates were allowed to cool in the deposition atmosphere to 100 $^{\circ}$C before being transferred to the load lock and removed from the system.
We determine the RuO$_2$ film thickness to be $\sim$155 nm using scanning electron microscsopy (SEM) measurements.

The optical transmittance and reflectance were measured from the UV to near IR (200 nm to 2200 nm)
at near-normal incidence (6 degrees off the surface normal) with a Cary UMA spectrophotometer.
Two small gaps in the data near 720 nm and 1100 nm occur due to grating and filter changes in the measurement.
A J.A. Woolam VASE ellipsometer collected data from 250 nm to 1100 nm at angles from 55 degrees to 80 degrees on
the 155 nm RuO$_2$ film on the TiO$_2$ (001) substrate.
The Raman scattering was measured in a backscattering geometry using 532 nm and 455 excitation wavelength sources.
Temperature dependent Raman spectra were recorded from 293 K to 403 K using 532 nm excitation
and the Raman frequencies were analyzed by fitting results with Gaussian functions.

\subsection{First-principles calculations}
We performed first-principles calculations
based on density functional theory \cite{hohenberg1964inhomogeneous,kohn1965self} within the
projector-augmented wave (PAW) method \cite{blochl1994projector} as
implemented in the Vienna Ab-initio Simulation Package
(VASP) \cite{kresse1993ab,kresse1994ab}.
All of the results in the main text use
the generalized gradient approximation (GGA) defined
by the Perdew-Burke-Ernzerhof functional \cite{perdew1996generalized}.
Since the GGA functional leads to a NM ground state we follow the approach adopted in previous
studies of altermagnetism in RuO$_2$ by applying a Hubbard-$U$ to the Ru $d$-states to obtain a magnetic moment of 1 $\mu_B$ on the Ru ions \cite{smejkal2022beyond}.
Using
the spherically averaged and rotationally invariant LDA+$U$ methodology proposed by Dudarev {\it et al.} \cite{dudarev1998electron} we find that a 
$U$ value of 1.6 eV leads to a moment of 1 $\mu_B$ on the Ru ions (obtained by considering the
Wigner-Seitz sphere when integrating the volume of charge density).

We use the VASP 5.4.4 Ru$_{\rm pv}$ and standard O PBE PAWs for all of the
calculations and a plane-wave energy cutoff of 600 eV.
We compared the results of our NM calculations obtained with calculations that use
the Ru$_{\rm sv}$ PAW and the GW PAWs and found
the electronic and optical properties to be approximately the same as those reported
in the main text.
The electronic density of states (DOS), vibrational properties and the
frequency dependent dielectric function \cite{gajdovs2006linear} were calculated
with a 38$\times$38$\times$44 $k$-point grid which was necessary to achieve converged results.
For the calculation of optical properties we verified the frequency dependent dielectric function is converged using a large number of empty states.  
For the DOS calculation we used the tetrahedron method while for the
calculation of the dielectric function we used Gaussian smearing with a broadening of 0.01 eV.
Spin-orbit coupling was included for the calculations of the DOS while
we used collinear calculations for the calculations of the frequency dependent dielectric function.
We verified the optical properties do not change with the inclusion of spin-orbit coupling (ESI \dag).

Optical properties were calculated using the density-density response of the frequency
dependent dielectric function.
The real part of the dielectric function, $\epsilon_r$ is obtained from the imaginary part of the
frequency-dependent dielectric function, $\epsilon_i$ using a complex shift of 0.1 eV.
We set the Drude relaxation rate to 0.2 eV for the NM and AM calculations of the frequency
dependent dielectric function.
To determine the optical properties such as the reflectivity, $R$, we use the imaginary and real
part of the frequency dependent dielectric function and calculated $R$ as follows:
\begin{equation}
R(\omega) = \frac{(n(\omega) - 1)^2 + \kappa(\omega)^2}{(n(\omega) + 1)^2 + \kappa(\omega)^2},
\end{equation}
where $n$ and $\kappa$ correspond to the complex refractive index and are defined as follows:
\begin{equation}
n = \sqrt{\frac{|\varepsilon_r + i \varepsilon_i| + \varepsilon_r}{2}}
\quad \text{and} \quad
\kappa = \sqrt{\frac{|\varepsilon_r + i \varepsilon_i| - \varepsilon_r}{2}}.
\end{equation}

To investigate the role of strain, the RuO$_2$ lattice parameters are strained to match the in-plane TiO$_2$ (001)
and (110) lattice parameters.
The phonon calculations are performed using Phonopy \cite{togo2015first}.

\section{Results and Discussion}
\subsection{Bulk properties}
We first discuss the results of our first-principles calculations on the structural, electronic and vibrational properties and compare them to
previous experimental measurements obtained on bulk RuO$_2$.
While there are prior reports on the electronic structure \cite{smejkal2022beyond} and vibrational properties \cite{Basak_Ptok_2024}
of RuO$_2$ in the AM state, a detailed comparison
between first-principles calculations of the NM versus the AM state
is scant \cite{wenzel2024fermi,smolyanyuk2024fragility}.
The NM and AM configurations relax to the rutile structure with the centrosymmetric P4$_2$/$mnm$
space group.
\begin{figure}[!h]
\includegraphics[width=8.5cm]{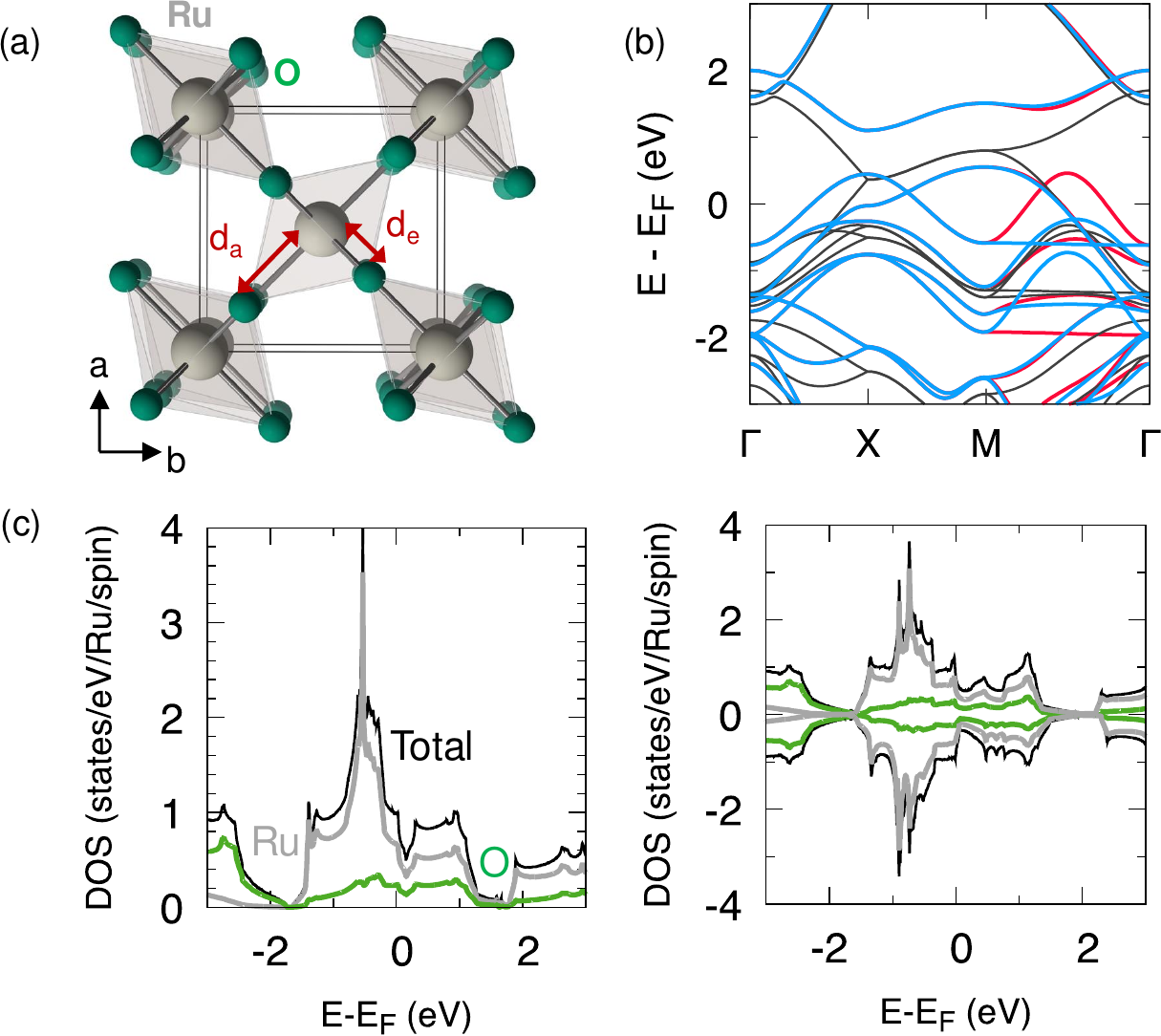}
\caption{
(a) Atomic structure of the RuO$_2$ unit cell.
The apical Ru-O bonds, $d_a$, and the equatorial Ru-O bonds, $d_e$, within the distorted octahedron are illustrated
with red arrows.
(b) Band structure of NM (black) and AM (majority spin in red and minority spin in blue)
RuO$_2$.
(c)  Atom-resolved density of states (DOS) of NM RuO$_2$ (left) and AM RuO$_2$ (right).  The
total density of states is illustrated in black, contributions from Ru in grey and O in green.
The band structure and DOS in (b) and (c) are plotted with respect to the Fermi level.}
\label{fig:ek_dos}%
\end{figure}
The Ru ions reside within distorted octahedra and are six-fold coordinated
by the O ions (Fig.~\ref{fig:ek_dos}(a)).  In the NM state the bond lengths of the two
apical Ru-O bonds is
1.951~\AA~and the four equatorial Ru-O bonds are 1.996~\AA, while in the AM state the apical Ru-O bond lengths are
1.950~\AA~and the equatorial Ru-O bond lengths are 2.010~\AA.
The ratio of the apical to equatorial Ru-O bond lengths is approximately the
same, $\sim$0.97, for the NM and the AM state.  The calculated lattice parameters for the AM and the NM
state are within $\sim$1\% of the experimentally measured lattice
parameters (Table \ref{table:vib}) of bulk RuO$_2$.
\begin{table*}[!htb]
\begin{center}
\setlength{\tabcolsep}{6pt} 
\renewcommand{\arraystretch}{1.4} 
\begin{tabular}{c|cccccccc}
  \toprule\toprule
& \makecell{NM \\ ($U$=0 eV)} 
  & \makecell{AM \\ ($U$=1.6 eV)} 
  & \makecell{NM \\ ($U$=1.6 eV)} 
  & \makecell{NM (001) \\ ($U$=0 eV)} 
  & \makecell{AM (001) \\ ($U$=1.6 eV)} 
  & \makecell{AM (110) \\ ($U$=0 eV)} 
  & \makecell{Experiment} \\
  \midrule
 \multicolumn{7}{c}{Lattice parameters (\AA)} \\
 \midrule
 a & 4.511 & 4.539 & 4.512 & 4.618 & 4.618 & 6.531,2.955 & 4.491 \\
 c & 3.131 & 3.132 & 3.137 & 3.071 & 3.095 & 6.475 & 3.106 \\
 \midrule
  \multicolumn{7}{c}{Density of states (states eV$^{-1}$ Ru$^{-1}$ spin$^{-1}$) }\\
  \midrule
  $N(E_{F})$ & 0.99 & 0.79 & 0.63 & 1.12 & 0.76 & 1.21 & 1.10 \\
  \midrule
   \multicolumn{7}{c}{Plasma frequencies (eV)} \\
   \midrule
  $\omega_{x}$ & 3.21 & 2.35 & 3.37 & 3.78 & 2.15 & 3.96,3.49 & 3.16 \\
  $\omega_{z}$ & 3.40 & 2.75 & 3.09 & 3.36 & 2.10 & 3.72 & 3.34 \\
\midrule
 \multicolumn{7}{c}{Raman frequencies (meV)} \\
 \midrule
  B$_{2g}$ & 88 & 90 & 88 & 79  & 81 & - & 87 \\
  A$_{1g}$ & 78 & 84 & 80 & 72  & 73 & - & 80 \\
  E$_g$ & 63 & 67 & 64 & 61  & 62 & - & 65 \\
  B$_{1g}$ & 20 & 19 & 20 & 17  & 22 & - & 20 \\
   \bottomrule\bottomrule
\end{tabular}
\end{center}
\caption{
First-principles calculations of the lattice parameters,
density of states at the Fermi level ($N(E_{F})$), plasma frequency ($\omega$), and the
frequency of the Raman active modes of RuO$_2$ in the
NM state, AM state, NM state using a Hubbard-$U$ of 1.6 eV, NM state strained to TiO$_2$(001) (NM (001)), AM state strained to TiO$_2$(001) (AM 001)) and the AM state strained to TiO$_2$(110) (AM (110)).  For each configuration we list the corresponding $U$ value that was used in the first row
of the table.  For AM (110) we list the two inequivalent $a$ and $b$ lattice constants of the orthorhombic cell within the row for $a$.
We compare our calculations with prior experiments on bulk RuO$_2$ \cite{rogers1969crystal,rosenblum1997raman,mertig1986specific,wenzel2024fermi}.}
\label{table:vib}
\end{table*}

The band structure and DOS in the NM and AM state are illustrated in Fig.~\ref{fig:ek_dos}(b-c).  The finite spin-splitting
that is characteristic of the AM state occurs along the $\Gamma$-M line of the Brillouin zone (Fig.~\ref{fig:ek_dos}(c)).
The DOS at the Fermi level, $N(E_{F})$, in the NM and AM state are comprised primarily
of Ru $4d$ orbitals with a minor admixture of O $2p$ states.
The magnitude of $N(E_{F})$ in the AM state is lower than that in the NM state (Table \ref{table:vib}).
The value of $N(E_{F})$
obtained from the Sommerfeld coefficient of the low temperature specific heat measurements \cite{mertig1986specific} is
1.10 states eV$^{-1}$ Ru$^{-1}$ spin$^{-1}$, which is only
slightly enhanced compared to the NM value for $N(E_{F})$, which is indicative of weak correlation effects.

The different values of $N(E_{F})$ of the NM versus the AM states leads to different plasma frequencies (Table \ref{table:vib}).
Since RuO$_2$ has a tetragonal crystal structure, there are two independent components to
the plasma frequency tensor - $\omega(x,y)$ which
corresponds to the crystallographic $a$ or $b$ directions and $\omega(z)$
which corresponds to the crystallographic $c$ direction.
The calculated plasma frequencies in the NM state are consistent with prior calculations \cite{de2006electronic,krasovska1995ab} and are
approximately similar to the experimentally
reported plasma frequencies.  The plasma frequency in the AM state is significantly lower compared to experiment.

While the degree of tetragonality of the rutile structure defined as $c/a$ is $\sim$0.6 for RuO$_2$
(where $a$ and $c$ are the lattice parameters from the NM DFT calculation) the electronic properties are
relatively isotropic if we consider the ratio $\omega_z/\omega_x$.
In the NM state $\omega_z/\omega_x$ is $\sim$1.06, which is similar to the ratio
obtained from experiment.
The isotropy of the $\omega_z/\omega_x$ ratio is also consistent with
direction-dependent electrical
transport measurements conducted on single crystal RuO$_2$ \cite{ryden1968temperature,kiefer2024crystal}
where they find relatively isotropic transport along the (100) and (001) directions across a wide range of temperatures.
In the AM state the ratio $\omega_z/\omega_x$ is slightly larger, $\sim$1.17, compared to the NM state, which
suggests a greater degree of anisotropy.

Rutile RuO$_2$ has four Raman-active modes labeled
by the following symmetries - B$_{1g}$, E$_g$, A$_{1g}$, and B$_{2g}$ - each involving only the displacement of O ions.  
Our NM calculations
lead E$_g$ and A$_{1g}$ modes that are 2 meV lower than experiment
while the B$_{2g}$ mode is 1 meV larger than experiment (Table \ref{table:vib}).
In the AM state, the A$_{1g}$, E$_g$ and the B$_{2g}$ modes involve displacements of Ru-O bonds in which the Ru ions are ferromagnetically
aligned.  The frequency of these three modes are
are up to 4 meV higher than the measured values and up to 6 meV larger than the frequencies in the NM state (Table \ref{table:vib}).

An additional consideration when comparing the calculated Raman frequencies with experiment is that
the DFT calculations correspond to $T$=0 K calculations while
the Raman measurements are at finite temperature.
If we use the experimental lattice parameters of bulk RuO$_2$ (Table \ref{table:vib}), the Raman frequencies
of the NM state shift by 1 meV relative to those calculated using the equilibrium structure while
in the AM state the frequencies decrease by 2 meV (ESI \dag).

One of the challenges with investigating altermagnetism in RuO$_2$ using conventional static DFT is that it relies on a large $U$ applied to the Ru states.  While this leads to a finite magnetic moment on the Ru ions, it also renormalizes the dispersion of the bands.  
To isolate the impact of applying $U$ on the predicted structural, electronic and vibrational properties, independent of the magnetic order,
we also optimized the RuO$_2$ structure and calculated $N(E_F)$, $\omega$ and the Raman modes frequencies in the NM state using $U$=1.6 eV (Table \ref{table:vib}).  The largest change with the inclusion of $U$ is a reduction of
$N(E_{F})$ by $\sim$30\% and an increase in the frequency of the $A_{1g}$ mode by 2 meV compared to the NM calculations with $U$=0 eV.

This suggests the underestimated
values of $N(E_{F})$ and $\omega$ in the AM state based on calculations using a Hubbard-$U$ compared to experiment are due to renormalization of the electronic structure that occurs by applying a Hubbard-$U$ to the
Ru states and the momentum dependent spin-splitting of the bands in the vicinity of the Fermi level in the AM state.
However, a $U$ value of 1.6 eV may be too large for a metallic system such as RuO$_2$
given the short-range screening in metals \cite{smolyanyuk2024fragility}.  Therefore, all of the NM results are analyzed with $U$=0 eV.

\subsection{Straining RuO$_2$ to TiO$_2$ (001) and (110)}

\begin{figure}[!htb]
\includegraphics[width=8.5cm]{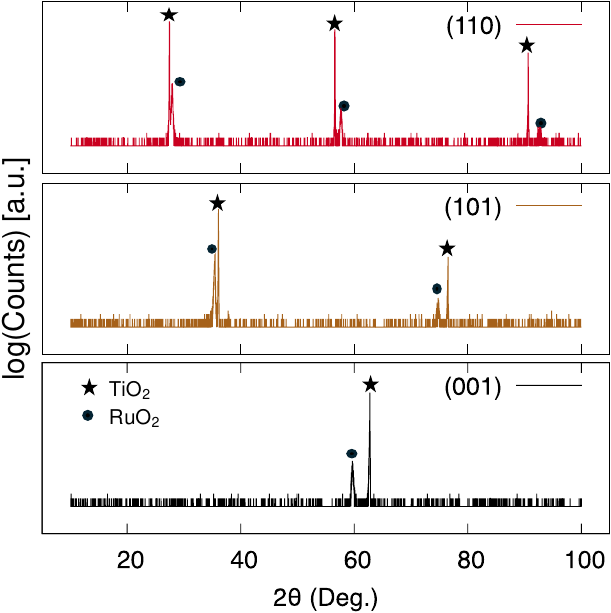}
\caption{
X-ray diffraction patterns collected on RuO$_2$ films grown on (001), (101), and (110)
TiO$_2$ substrates.}
\label{fig:xrd}%
\end{figure}

Next we consider the impact that strain has on the properties of RuO$_2$.
We grew RuO$_2$ films on (001), (101), and (110) TiO$_2$ substrates.
X-ray diffraction (XRD) patterns collected on each of these films (Fig.~\ref{fig:xrd}) lead to peaks that correspond
only to the out-of-plane reflections, which is indicative of the heteroepitaxial relationship
between the RuO$_2$ films and the TiO$_2$ substrates \cite{fields2024orientation}.
The RuO$_2$ XRD peaks are shifted compared to their expected bulk values
due to a combination of lattice strain and differences in the thermal expansion coefficients
of RuO$_2$ and TiO$_2$.

To quantify the impact of this strain, we calculated the structural, electronic, and vibrational properties of RuO$_2$ with the in-plane lattice constants constrained to match TiO$_2$ (001) and (110) while allowing the RuO$_2$ $c$-lattice constant and all atomic positions to relax.
For the (001) orientation, we fix the in-plane lattice constants of RuO$_2$ to 4.618 Å, matching the TiO$_2$ lattice constant obtained from DFT. The 2.3\% tensile in-plane strain compresses the RuO$_2$ $c$-axis, which is accompanied by an overall increase in unit cell volume.  The space group of the strained unitcell remains P4$_2$/$mnm$.  Under this strain, both $N(E_F)$ and $\omega$ increase by up to 10\%~(Table \ref{table:vib}) while the Raman frequencies decrease by up to 9 meV compared to the equilibrium NM state.   We were unable to stabilize any magnetic moment on the Ru ions in the AM state using GGA calculations.
Applying a Hubbard-$U$ of 1.6 eV to the strained RuO$_2$ (001) cell, stabilizes the AM state over the NM state by 55 meV per Ru ion and leads to a magnetic moment of 1 $\mu_B$ on the Ru ions. It also leads to spin-splitting of the bands along the $\Gamma$–M line of the Brillouin zone, similar to what we find using the equilibrium lattice parameters of RuO$_2$ (Fig.~\ref{fig:ek_dos}(b)).
Notably, a lower $U$ value of 1.2 eV is sufficient to 
stabilize a moment of 1 $\mu_B$ on the Ru ions in the strained RuO$_2$ (001) cell.
We also checked whether the ferromagnetic (FM) state is stable in the strained RuO$_2$ (001) cell.
Similar to the case of equilibrium RuO$_2$, we find ferromagnetic order is not stable with the magnetic moment on each Ru ion converging close to zero in the calculations with and without a Hubbard-$U$. 

\begin{figure}[!h]
\includegraphics[width=8.5cm]{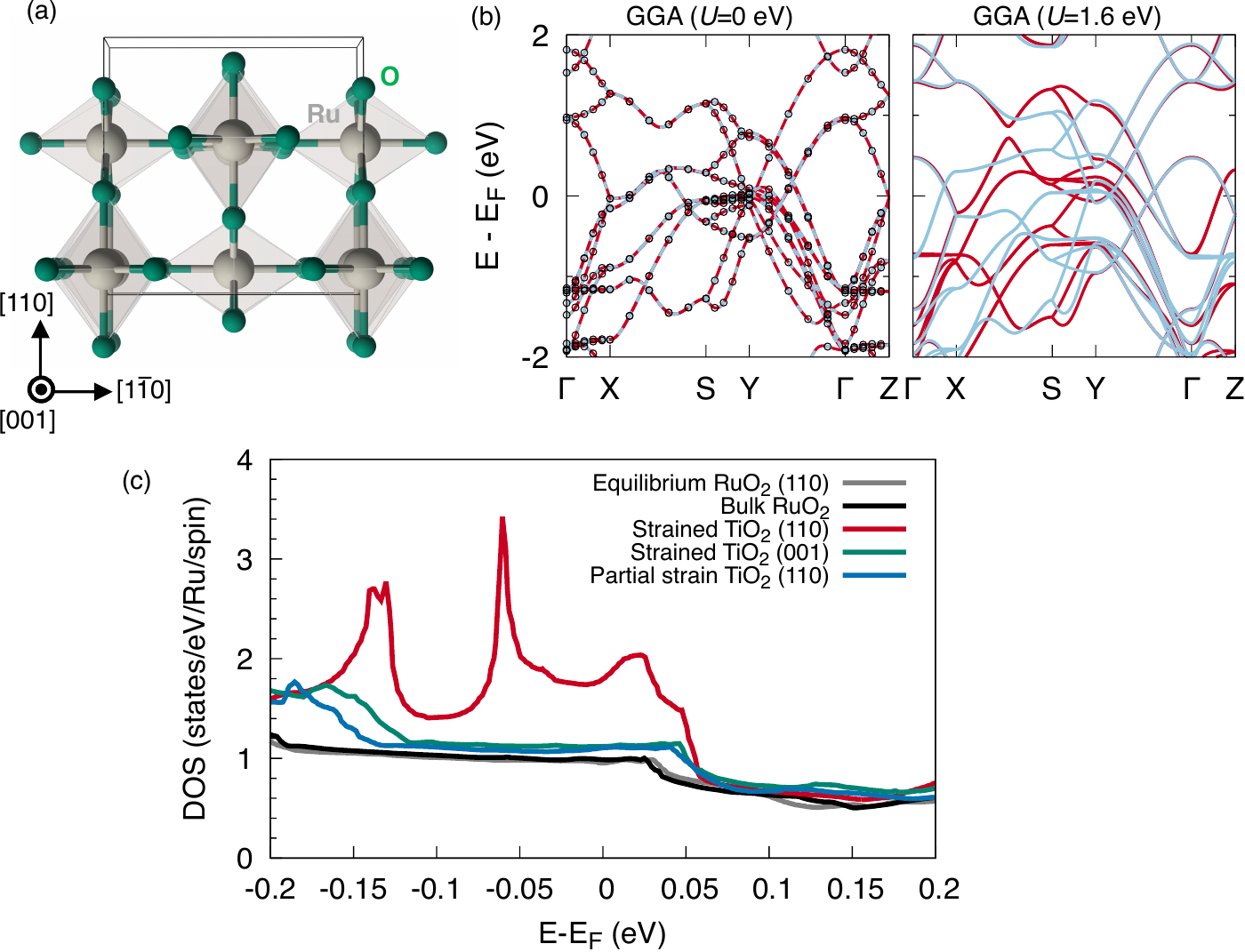}
\caption{
(a) Atomic structure of the orthorhombic RuO$_2$ (110) unit cell.
(b) Band structure of RuO$_2$ (110) strained to TiO$_2$ (110) 
obtained with GGA without a Hubbard-$U$ (left) and with $U$=1.6 eV (right).  The GGA calculations illustrate the NM state (black circles) alongside AM state (majority spin in red and minority spin in blue) while the calculations with a finite $U$ only illustrate the AM state.  GGA calculations in the FM state also leads to a negligible spin splitting (not shown) with respect to the NM calculations.
(c) Non-magnetic GGA density of states for equilibrium RuO$_2$ (110) (grey), bulk RuO$_2$ (black), RuO$_2$ strained to TiO$_2$ (110) (red), RuO$_2$ strained to TiO$_2$ (001) (green) and RuO$_2$ partially strained to TiO$_2$ (110) (blue).  See the main text for the lattice constants used for each calculation.
}
\label{fig:ek_110}%
\end{figure}

For RuO$_2$ strained to TiO$_2$ (110), we transform the tetragonal rutile unitcell to an orthorhombic cell (space group $Pmm2$) where the in-plane $a$ and $b$ lattice vectors correspond to the [1$\bar{1}$0] and [001] directions (Fig.~\ref{fig:ek_110}(a)), respectively, of the RuO$_2$ rutile unit cell and the out-of-plane $c$-lattice constant is the [110] direction of the rutile unitcell.  The orthorhombic unit cell leads to two pairs of inequivalent Ru ions with slightly different Ru-O bond lengths between each pair.  Optimizing the orthorhombic structure using NM GGA calculations leads to $a$=6.380 \AA~and $b$=3.131 \AA.  To simulate epitaxial strain to TiO$_2$ (110) we fix $a$ to 6.531 \AA~and $b$ to 2.955 \AA~, which are the
[$\bar{1}$10] and [001] DFT lattice constants of TiO$_2$, which amounts to 2.4\%~tensile and 5.6\%~compressive strain, respectively.  This anisotropic in-plane strain imposed by TiO$_2$ (110) increases the RuO$_2$ $c$-lattice constant by 3.4\% (Table \ref{table:vib}).

In the strained RuO$_2$ (110) cell GGA calculations, the FM state is 4.4 meV per Ru ion lower in energy compared to the NM state while the AM state is 3.5 meV per Ru ion lower in energy than the NM state.   In the AM state the magnetic moment is 0.3 $\mu_B$ on each of the Ru ions while in the FM state there are two different ferromagnetically aligned moments, 0.1 $\mu_B$ on one pair of Ru ions and 0.3 $\mu_B$ on the second pair.  Within LDA the magnetic moment is reduced to 0.1 $\mu_B$ in the AM state and 0.08 $\mu_B$ and 0.2 $\mu_B$ on the two pairs of Ru ions in the FM state, which reflects the sensitivity of the magnetic moment to the exchange correlation approximation.  Consistent with the FM state being lower in energy, $N(E_{F})$ in the FM state (1.45 states eV$^{-1}$ Ru$^{-1}$ in the majority spin and 0.70 states eV$^{-1}$ Ru$^{-1}$ in the minority spin) and the AM state (1.21 states eV$^{-1}$ Ru$^{-1}$)is lower than the value of $N(E_{F})$ in the NM state (1.81 states eV$^{-1}$ Ru$^{-1}$) based on GGA calculations. 
The low magnetic moment in the AM and FM states leads to negligible spin-splitting of the bands in comparison to calculations of the electronic structure in the NM state (Fig.~\ref{fig:ek_110}(b)).

When a Hubbard-$U$ of 1.6 eV is applied to the RuO$_2$ (110) strained cell, the AM state is 31 meV per Ru ion lower in energy than the NM state while the FM state can not be stabilized.  The magnetic moment on the Ru ions in the AM state is 1 $\mu_B$ state which leads to a larger momentum dependent spin-splitting of the band structure (Fig.~\ref{fig:ek_110}(b)).  The combination of Hubbard-$U$ and spin-splitting leads to a lower $N(E_{F})$ is 0.79 states eV$^{-1}$ Ru$^{-1}$ in the AM ground state relative to $N(E_{F})$ obtained without applying a finite $U$.  

Straining RuO$_2$ to match the TiO$_2$ (110) substrate appears to be essential for stabilizing the FM and AM states. 
When using the relaxed RuO$_2$ (110) lattice parameters, the NM state remains the ground state, and the AM and FM states cannot be stabilized. This remains true even under partial strain conditions, where the in-plane lattice constants of RuO$_2$ are set to 6.436~\AA\ and 3.033~\AA, 
corresponding to 0.9\%~tensile strain along [1$\bar{1}$0] and 3.1\%~compressive strain along [001] relative to the equilibrium values. 
We also note that when RuO$_2$ is fully strained to match TiO$_2$ (110) $N(E_F)$ in the nonmagnetic state increases 
significantly compared to all other cases considered, including bulk RuO$_2$, equilibrium RuO$_2$ (110), RuO$_2$ partially 
strained to TiO$_2$ (110), and RuO$_2$ strained to TiO$_2$ (001) (Fig.~\ref{fig:ek_110}(c)). 
This enhanced $N(E_F)$ in the nonmagnetic state under full (110) strain may be one possible reason for the tendency to magnetism
in this configuration.

Hence, we suggest that RuO$_2$ strained to TiO$_2$ (001) is likely non-magnetic since the AM state is only stabilized if we use a Hubbard-$U$ greater than 1 eV, which is relatively large for a metallic system.  When RuO$_2$ is completely strained to TiO$_2$ (110) this leads to the FM and AM states being nearly degenerate in energy and lower in energy relative to the NM state, consistent with prior first-principles studies  \cite{jeong2025metallicity}.  The energy difference between the FM and AM states in the strained RuO$_2$ cell GGA calculations (0.9 meV per Ru) is within the precision limit of our first-principles calculations and exhibits a great degree of sensitivity to the amount of strain and the exchange-correlation approximation used in the calculations.

\subsection{Raman and optical spectroscopy: experiments and comparisons with first-principles calculations}
\begin{figure}[!htb]
\includegraphics[width=8.5cm]{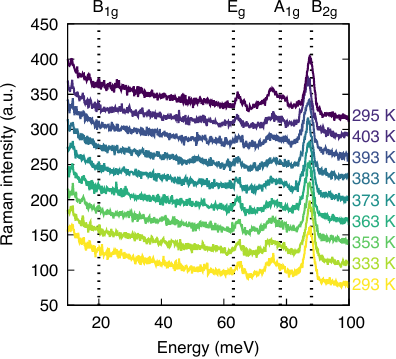}
\caption{
Temperature-dependent Raman scattering measurements using 532 nm excitation
for RuO$_2$ grown on (001) TiO$_2$
with the measurement temperature listed to the right of each spectra.
The spectra at each temperature are offset vertically for clarity.
The calculated Raman frequencies for NM RuO$_2$ are marked with dotted lines and the corresponding modes listed at the top.
}
\label{fig:ramanT}%
\end{figure}

We now turn to our
measurements of the optical properties and Raman scattering of the RuO$_2$ films grown on TiO$_2$.
Based on our unpolarized and polarized Raman spectroscopy measurements we identify the E$_g$, A$_{1g}$, and B$_{2g}$ Raman modes of rutile RuO$_2$ for the films grown on TiO$_2$ (001), (110) and (101). The Raman mode frequencies show a weak dependence on substrate orientation and deviate from the bulk values by up to 2 meV (ESI \dag), likely due to partial relaxation of the films \cite{fields2024orientation}.
We predict that a transition from the NM to the AM state of rutile RuO$_2$ leads to a softening of the A$_{1g}$ and E$_g$ Raman mode frequencies by up to 6 meV (Table \ref{table:vib}). Motivated by recent reports of a N$\acute{\rm e}$el transition in RuO$_2$ films near 350 K (77 $^{\circ}$C) \cite{jeong2024altermagnetic,song2025spin}, we investigate whether temperature-dependent Raman measurements of the RuO$_2$ films reveals changes in the frequencies of the Raman modes (Fig.~\ref{fig:ramanT}). Increasing the temperature to 403 K softens the frequencies of the  the B$_{2g}$, A$_{1g}$, and E$_g$ modes by approximately 0.9 meV and decreases the intensity of the A${1g}$ and E$g$ modes.  The Raman modes associated with the TiO$_2$ substrate exhibited a comparable degree of softening.
Taken together, the relatively small magnitude of the frequency shifts and the absence of any abrupt changes in the spectra do not support the presence of a magnetic phase transition in RuO$_2$ over the temperature range investigated.

\begin{figure}[!h]
\includegraphics[width=8.5cm]{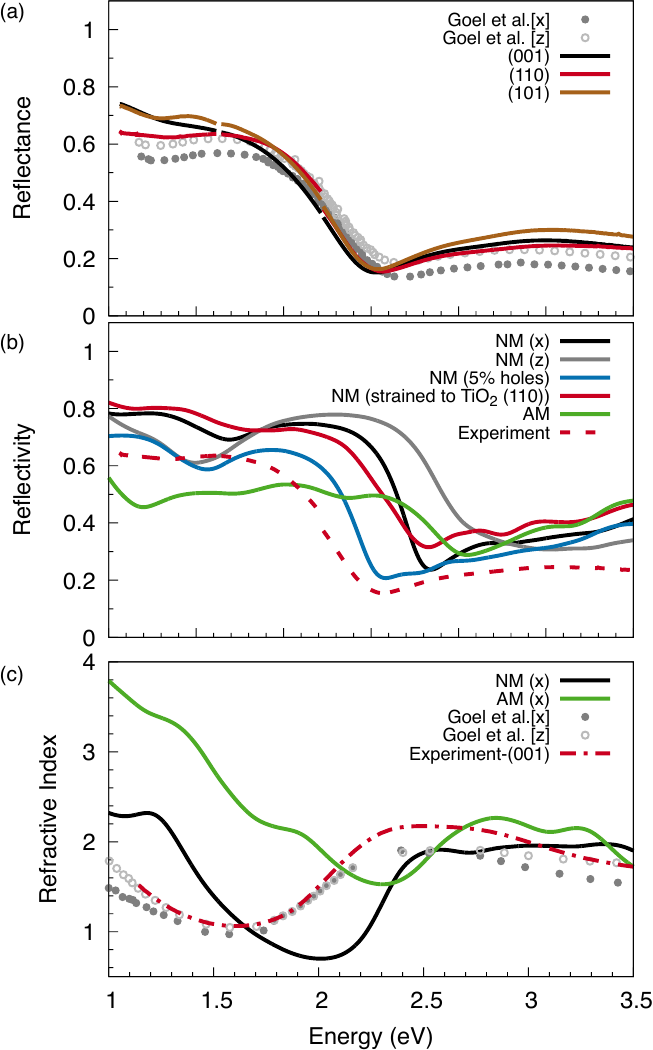}
\caption{
(a) Experimental measurement of optical reflectance of RuO$_2$ grown on TiO$_2$ (001) (black), (101) (orange) and (110) (red)
substrates compared to
experiments conducted on bulk single crystal RuO$_2$ along the in-plane direction (filled grey circles) and out-of-plane
direction (open grey circles)
\cite{goel1981optical}.
(b) Optical reflectivity obtained from first-principles calculations of the NM state along the
in-plane (x) and out-of-plane (z)
directions of the RuO$_2$ crystal, AM state, NM state strained to RuO$_2$ (110) and RuO$_2$ doped with 5\%~holes.
Optical reflectance measured in RuO$_2$ grown on TiO$_2$ (110) (dotted lines).
(c) Real part of the refractive index calculated for NM RuO$_2$, AM RuO$_2$ and compared against measurements on bulk RuO$_2$ \cite{goel1981optical} and our measurements of RuO$_2$ on TiO$_2$ (001).
Calculations along the in-plane direction are marked with (x) and along the out-of-plane
direction are marked with (z) in the legend.
The experimental data \cite{goel1981optical} in panels (a) and (c) was reproduced with permission
from Phys. Rev. B 24, 7342 (1981). Copyright 1981 American Physics Society.
}
\label{fig:reflectivity}%
\end{figure}

Next we measured the optical properties of our RuO$_2$ films motivated in
part by the proposal that magneto-optical measurements
\cite{zhou2021crystal,weber2024all,rao2024tunable} have
been suggested as a plausible route to detect AM order.
Our measurements of the optical reflectance of the RuO$_2$ films 
are in agreement with measurements reported on bulk RuO$_2$ \cite{goel1981optical}
both in terms of the magnitude and spectral dependence (Fig.~\ref{fig:reflectivity}(a)).
The experimentally measured reflectance
decreases for energies greater than $\sim$1 eV and reaches a minimum $\sim$2.2. Above 2.2 eV, the reflectance
is relatively independent of the photon energy.  The spectral dependence of the reflectance in our measurements is
relatively independent of the TiO$_2$ orientation which as we will show is due to the weak effect of strain imposed by TiO$_2$ on the optical reflectivity.

In Fig.~\ref{fig:reflectivity}(b) we compare our experimentally measured reflectance
with our first-principles calculations of reflectivity in the NM and AM state.
Our NM calculations captures the overall magnitude and spectral dependence of the reflectance.
When RuO$_2$ is strained to the TiO$_2$ (110) and TiO$_2$ (001) (not shown), the dip in the reflectivity shifts to slightly lower energies by $\sim$ 100 meV while the overall spectral dependence is similar to the reflectivity obtained with the unstrained lattice parameters of the NM state.
Comparing the calculated reflectivity in the NM state with the DOS calculations (Fig. \ref{fig:ek_dos}) we suggest
the dip in reflectivity at $\sim$2 eV, either corresponds to optical transitions between the O-$2p$ states
below the Fermi level and the unoccupied Ru $d$-states above the Fermi level or transitions between occupied Ru $d$-states at the Fermi level and the antibonding states at higher energies  (Fig.~\ref{fig:reflectivity}).

Calculations in the AM state leads to optical reflectivity that is different from experiment.
Between 0.5 eV and 2 eV, the reflectivity in the AM state is $\sim$30\%~lower than the reflectance measured in experiment and the
noticeable reduction in the reflectivity at $\sim$ 2 eV seen in experiment and our NM calculations is absent.
For energies between $\sim$ 2.5 and 3.5 eV,
the difference in the magnitude of the reflectivity between the AM state and NM state
is less, differing by up to $\sim$ 20\%.  We surmise the greater discrepancy of the reflectivity in the AM state compared to the NM
state and to experiment
for energies below $\sim$ 2.5 eV is due to interband transitions that lead to larger absorption (lower reflectivity) involving
the spin-split bands of the AM state.

Finally, we assess whether hole doping affects the reflectivity given that the presence of ruthenium vacancies
have been
raised as a possible origin of the AM reported in the prior experimental studies \cite{smolyanyuk2024fragility}.  We consider
a hole doping concentration of 5\%~within the rigid-band approximation.  For this doping concentration we find
there is no
change in the spectral dependence of the reflectivity compared to the
NM state (Fig.~\ref{fig:reflectivity}(b)).

To connect our optical reflectance measurements more directly with the electronic structure of RuO$_2$
we also measured the refractive index, focusing on RuO$_2$ grown on TiO$_2$ (001) (Fig.~\ref{fig:reflectivity}(c)).
The measured real part of the refractive index, $n$, agrees with prior measurements on bulk RuO$_2$ \cite{goel1981optical} (Fig.~\ref{fig:reflectivity}(c)).  Both experiments
exhibit a pronounced dip in $n$ at $\sim$1.6 eV.  For photon energies above $\sim$1.85 eV, the spectral dependence of the optical properties is primarily determined by $n$, since its magnitude exceeds that of the imaginary component, $k$, in this range (ESI \dag).
We also present calculations of $n$ for both the NM and AM states. In the NM calculations, a dip in $n$ appears near 1.9 eV, and for higher energies, both the magnitude and spectral shape of $n$ agree well with the experimental data.
In contrast, the AM state exhibits a dip in $n$ at a significantly higher energy, around 2.3 eV. Below this energy, the magnitude of $n$ is substantially overestimated relative to experiment. However, for energies above 2.3 eV, the spectral dependence of $n$ in the AM state more closely matches experiment, likely due to the weaker AM-induced splitting at energies far from the Fermi level.

\section{Conclusions}
In conclusion, we have combined Raman and optical spectroscopy measurements of RuO$_2$ films grown on TiO$_2$ (001), (101), and (110) substrates with first-principles calculations of RuO$_2$ in  the nonmagnetic and altermagnetic state.  While our experiments do not offer direct insight into the presence or lack of time-reversal symmetry breaking, the comparisons between experiment and theory allows us to draw the following conclusions about RuO$_2$:
\begin{enumerate}
\item Nonmagnetic GGA calculations leads to electronic properties that are consistent with experiment, Raman frequencies that are within 2 meV of the measured values, and reproduces the magnitude and spectral dependence of the optical properties.
\item Altermagnetic GGA+$U$
calculations leads to Raman frequencies that are up to 4 meV higher
compared to experiment, and fails to describe the electronic and optical properties, both in terms of magnitude and spectral dependence.
\item RuO$_2$ strained to TiO$_2$ (001) remains nonmagnetic while straining to TiO$_2$ (110) leads to the altermagnetic and ferromagnetic states being near degenerate in energy and are both lower in energy than the nonmagnetic configuration based on GGA calculations.  Applying a large Hubbard-$U$ stabilizes the altermagnetic state for RuO$_2$ strained to TiO$_2$ (001) and (110)
\end{enumerate}
Based on this analysis, first-principles calculations of RuO$_2$ in the nonmagnetic state provide a consistent description of the optical and vibrational properties of bulk RuO$_2$.  RuO$_2$ strained to TiO$_2$ (001) remains nonmagnetic while straining to TiO$_2$ (110) stabilizes a magnetic state with a small moment on Ru, which highlights the sensitivity of magnetic ordering to epitaxial strain in RuO$_2$.

\section*{Conflicts of interest}
There are no conflicts to declare.

\section*{Data availability}
The data supporting this article have been included as part
of the (ESI \dag).  

\section*{Acknowledgements}
This work was supported by the Office of Naval Research through the Naval Research Laboratory's Basic Research Program.
Calculations were performed at the DoD Major Shared Resource Centers at AFRL and the Army ERDC.


%
\end{document}